\documentclass[12pt, a4paper]{article}
\usepackage{cmap}
\usepackage[T2A]{fontenc}		
\usepackage[utf8]{inputenc}		
\usepackage[ukrainian, english]{babel}	
\usepackage[colorlinks,hypertexnames=false]{hyperref}
\usepackage{cite}
\usepackage{graphicx}  
\graphicspath{{images/}{images2/}}  
\usepackage[justification=centering,singlelinecheck=false]{caption}

\usepackage{amssymb,amsfonts,amsmath,amsthm}

\usepackage[usenames]{color}
\usepackage{colortbl}
\usepackage[dvipsnames]{xcolor}

\usepackage{array,tabularx,tabulary,booktabs} 

\renewcommand{\phi}{\varphi}

\counterwithin{equation}{section}

\usepackage{indentfirst}

\usepackage[titletoc, title]{appendix}

\begin{document}
\selectlanguage{english}
\title{Divergences in the effective loop interaction of the Chern-Simons bosons with leptons. The unitary gauge case}
\author{Yuliia  Borysenkova, Volodymyr Gorkavenko${}^1$, Ivan Hrynchak${}^1$,\\ Oleksandr Khasai${}^2$,
Mariia Tsarenkova${}^1$ \vspace{1em}\\
${}^1$ \it \small Faculty of Physics, Taras Shevchenko National University of Kyiv,\\ 
\it \small 64, Volodymyrs'ka str., Kyiv 01601, Ukraine,\\
${}^2$ \it \small Bogolyubov Institute for Theoretical Physics, National Academy of Sciences of Ukraine,\\
\it \small 14-b Metrolohichna str., Kyiv 03143, Ukraine}
\date{}

\maketitle
\setcounter{equation}{0}
\setcounter{page}{1}%
    \begin{abstract}
     In this paper, we consider the extension of the Standard Model (SM) with Chern-Simons type interaction. This extension has a new vector massive boson (Chern-Simons bosons). There is no direct interaction between the Chern-Simons bosons and fermions of the SM.
       Using only three-particle dimension-4 interaction of the Chern-Simons bosons with vector bosons of the SM, we consider effective loop interaction of a new vector boson with leptons.  We consider the renormalizability of this loop interaction and conclude that
       for the computation of loop diagrams in the unitary gauge, we can not eliminate the divergences in the effective interaction of the Chern-Simons bosons with leptons. 
       
    \end{abstract}    
    \selectlanguage{english}

    \section{Introduction}

Despite the success of the Standard Model (SM) \cite{Cottingham:2007zz} in describing numerous collider experiments, SM is not a complete theory because it cannot explain phenomena such as active neutrino oscillation  (see e.g. \cite{Bilenky:1987ty,Strumia:2006db,deSalas:2017kay}), baryon asymmetry of the Universe (see e.g. \cite{Steigman:1976ev,Riotto:1999yt,Canetti:2012zc}),  dark matter (see e.g. \cite{Peebles:2013hla,Lukovic:2014vma,Bertone:2016nfn}). 
In all likelihood, the SM needs to be extended to include new particles and new interactions. 

It may turn out that SM has a hidden sector with many new particles, and some of these particles are not relevant to solving problems of the SM. We do not detect new particles because they are quite heavy or light but very feebly interact with particles of the SM.  If the new particles are quite heavy and can not be produced at current accelerators, we hope they will be detected at more powerful future accelerators such as FCC \cite{Golling:2016gvc,FCC:2018byv}. But if the new particles are light they can be found at existing accelerators nowadays, see e.g. \cite{Gorkavenko:2019nqm,Beacham:2020,Lanfranchi:2020crw}, in the intensity frontier experiments such as  MATHUSLA \cite{Curtin:2018mvb}, FACET \cite{Cerci:2021nlb}, FASER \cite{FASER:2018ceo,FASER:2018eoc}, SHiP \cite{Anelli:2015pba,Alekhin:2015byh}, NA62 \cite{Mermod:2017ceo,NA62:2017qcd,Drewes:2018gkc}, DUNE \cite{DUNE:2015lol,DUNE:2020fgq},  LHCb \cite{Gorkavenko:2023nbk}, etc.

We don’t know what type of particles of the new physics they will be. They can be scalar \cite{Patt:2006fw,Bezrukov:2009yw,Boiarska:2019jym}, pseudoscalar (axionlike) \cite{Peccei:1977hh, Weinberg:1977ma, Wilczek:1977pj,Choi:2020rgn}, fermion \cite{Asaka:2005pn,Asaka:2005an,Bondarenko:2018ptm,Boyarsky:2018tvu}, or vector (dark photons) \cite{Okun:1982xi,Holdom:1985ag,Langacker:2008yv} particles, see reviews \cite{Curtin:2018mvb,Alekhin:2015byh}. In this paper, we consider the extension of the Standard Model with Chern-Simons type interaction with a new massive vector boson (Chern-Simons bosons or CS bosons in the following). The Chern-Simons interactions appear in various theoretical models, including extra dimensions and string theory. 
see e.g. \cite{Antoniadis:2000ena,Coriano:2005own,Anastasopoulos:2006cz,Harvey:2007ca,Anastasopoulos:2008jt,Kumar:2007zza}.
The minimal gauge-invariant Lagrangian of the interaction of the CS bosons with SM particles has the form 
of 6-dimension operators \cite{Alekhin:2015byh,Antoniadis:2009ze}: 
\begin{align}
    \mathcal{L}_1&=\frac{C_Y}{\Lambda_Y^2}\cdot X_\mu (\mathfrak D_\nu H)^\dagger H B_{\lambda\rho} \cdot\epsilon^{\mu\nu\lambda\rho}+h.c.,\label{L1} \\
    \mathcal{L}_2&=\frac{C_{SU(2)}}{\Lambda_{SU(2)}^2}\cdot X_\mu (\mathfrak D_\nu H)^\dagger F_{\lambda\rho} H\cdot\epsilon^{\mu\nu\lambda\rho}+h.c.,\label{L2}  
\end{align}
where $\Lambda_Y$, $\Lambda_{SU(2)}$ are new scales of the theory; $C_Y$, $C_{SU(2)}$ are new dimensionless coupling constants;  $\epsilon^{\mu\nu\lambda\rho}$ is the Levi-Civita symbol ($\epsilon^{0123}=+1$); $X_\mu$ -- CS vector bosons; $H$ -- scalar field of the Higgs doublet; $B_{\mu\nu}=\partial_\mu B_\nu-\partial_\nu B_\mu$, $F_{\mu\nu}={\displaystyle -ig\sum_{i=1}^3\frac{\tau^i}{2} V^i_{\mu\nu} }$ -- field strength tensors of the $U_Y(1)$ and $SU_W(2)$ gauge fields of the SM. The gauge-invariant of the Lagrangians \eqref{L1}, \eqref{L2} is achieved because $X_\mu$ is the Stueckelberg field \cite{Ruegg:2003ps,Kribs:2022gri}.

After the electroweak symmetry breaking Lagrangians \eqref{L1}, \eqref{L2}  generate (among other terms
of higher dimensions) Lagrangian of three-field interactions
in  the  form  of 4-dimensional operators: 
\begin{equation}\label{Lcs}  
     \mathcal{L}_{CS}=c_z \epsilon^{\mu\nu\lambda\rho} X_\mu Z_\nu \partial_\lambda Z_\rho +c_\gamma \epsilon^{\mu\nu\lambda\rho} X_\mu Z_\nu \partial_\lambda A_\rho+\left\{ c_w \epsilon^{\mu\nu\lambda\rho} X_\mu W_\nu^- \partial_\lambda W_\rho^+ + h.c.\right\},
\end{equation}
where $A_\mu$ is the electromagnetic field; $W^\pm_\mu$ and $Z_\mu$ are  fields of the weak interaction; and $c_z$, $c_\gamma$, $c_w$ are some dimensionless independent coefficients. Coefficients $c_z$ and $c_\gamma$ are real, but $c_w$  can be complex. As one can see, there is no direct interaction of the CS vector boson $X_\mu$ with fermions of the SM.

 If one rewrites the coefficients before operators in \eqref{L1}, \eqref{L2}  as
 $C_Y/\Lambda_Y^2=C_1/v^2$ and $C_{SU(2)}/\Lambda_{SU(2)}^2=C_2/v^2$, where $C_1=c_1+{\rm i} c_{1{\rm i}}$ and  $C_2=c_2+{\rm i} c_{2{\rm i}}$ are dimensionless coefficients, then in unitary gauge one can obtain
\begin{align}
&    c_z=  -c_{1{\rm i}}g' + \frac{c_{2}}2 g^2, \label{L7}\\
&    c_\gamma = c_{1{\rm i}} g + \frac{c_{2}}2 g g^{'} , \label{L8}\\
&    c_w= \frac{c_2+{\rm i} c_{2{\rm i}}}2 g^2\equiv \Theta_{W1}+{\rm i} \Theta_{W2}.\label{L9}
\end{align}

Effective interaction of the CS bosons with quarks of different flavours was considered in \cite{Dror:2017ehi,Dror:2017nsg,Borysenkova:2021ydf}. It was shown that in this case divergent part of the loop diagrams (containing only $W^\pm$ bosons) is proportional to a non-diagonal element of the unity matrix $U^+U$ ($U$ is the Cabibbo–Kobayashi–Maskawa matrix) and is removed. It allows one to construct an effective Lagrangian of the interaction of the CS bosons with quarks of different flavours and compute the GeV-scale CS bosons' production in decays of meson. At the same time, it was shown that effective interaction of the CS bosons with quarks of the same flavours or with leptons contains divergence. This divergence can not be removed via counterterms of the CS boson interaction with fermions because the initial Lagrangian does not contain these terms.

The question of the effective interaction of the CS bosons with fermions of the same flavour is very important. Solving this problem will make it possible to find all decay channels of the CS boson and compute the CS boson’s lifetime, which in turn will allow us to compute the sensitivity region of the intensity frontier experiments for searching for the CS boson.

In this paper, we will consider the effective interaction of the CS bosons with leptons and compute all corresponding loop diagrams (containing $W^\pm$, $Z$ bosons, and photons) in the unitary gauge. We will be interested in the processes of decay of the CS boson into a lepton pair $\ell^+\ell^-$ ($\ell=e,\mu,\tau$) and neutrinos. 
We want to check whether there will be a cancellation of divergences when taking into account all corresponding diagrams.

    \section{Lepton pair production in the CS boson decays.\\ Direct computations}

\begin{figure}
    \centering
    \includegraphics[width=0.7\textwidth]{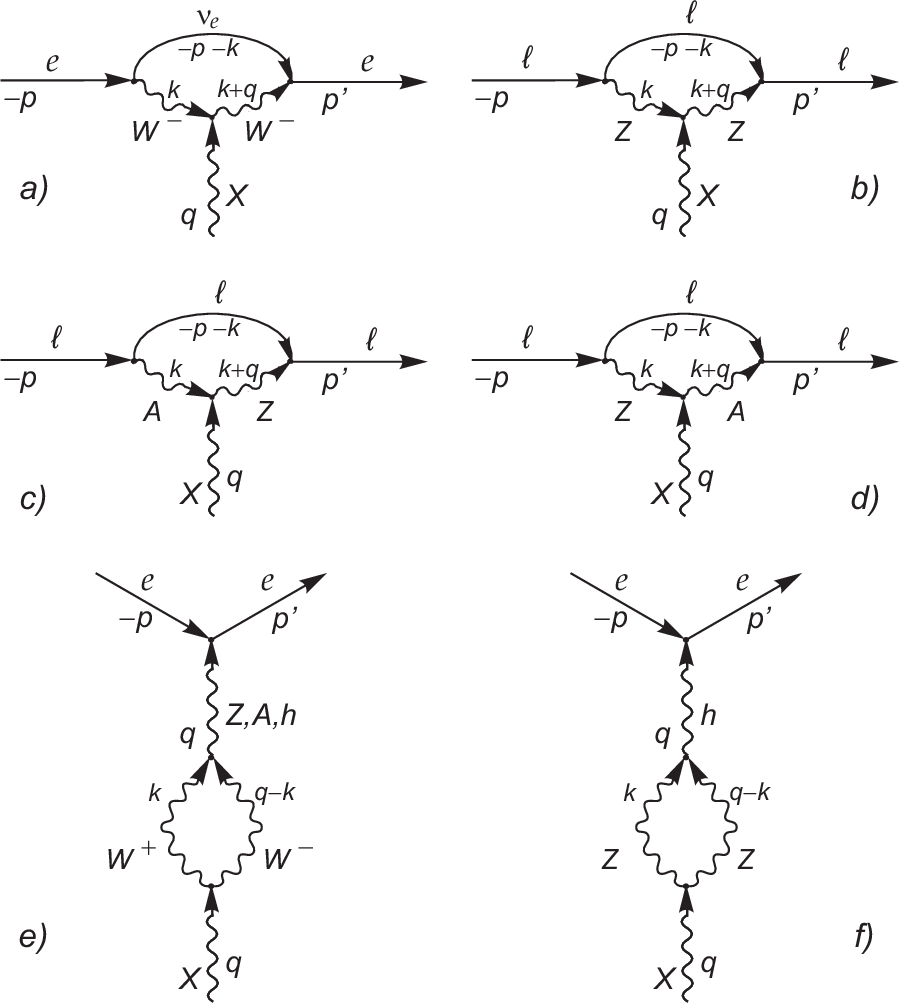}
    \caption{Diagrams of the CS boson's decay into leptons in the unitary gauge.}
    \label{fig:decay_leptons}
\end{figure}

\begin{figure}
    \centering
    \includegraphics[width=0.65\textwidth]{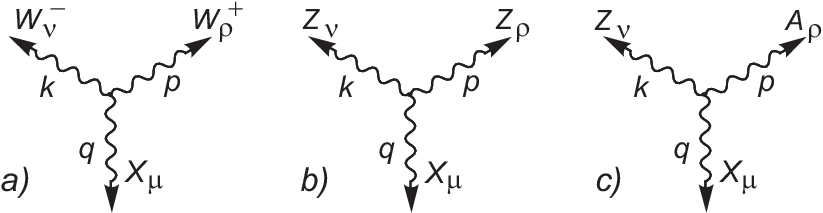}
    \caption{Vertex diagrams of the interaction of the CS boson with vector fields of the SM following \eqref{Lcs}.}
    \label{fig:vertex_diagrams}
\end{figure}

In the unitary gauge to compute the amplitude of the CS boson's decay into leptons one has to take into account the following diagrams presented in Fig.\ref{fig:decay_leptons}, where for the vertex of the CS boson interaction with vector fields \eqref{Lcs} we have the following rules, see Fig.\ref{fig:vertex_diagrams}
\begin{align}
    & XWW \qquad - (c_W p - c_W^* k)_\lambda \epsilon^{\mu\nu\lambda\rho},\\
    & XZZ \qquad -c_Z p_\lambda \epsilon^{\mu\nu\lambda\rho},\\
    & XZA \qquad -c_\gamma p_\lambda \epsilon^{\mu\nu\lambda\rho}.
\end{align}
It is also necessary to take into account that the diagram with vertex $XZZ$ in Fig.\ref{fig:decay_leptons}\textit{b} and Fig.\ref{fig:decay_leptons}\textit{f} actually corresponds to two diagrams (line with a derivative from $Z$ can be from the left side or the right side of the diagram).

\subsection{Production via $XWW$ interaction}

Lepton pair production via $XWW$ interaction is described by the diagrams \textit{a} and \textit{e} in  Fig.\ref{fig:decay_leptons}. 
The amplitude of
the lepton pair production via diagram \textit{e}  in   Fig.\ref{fig:decay_leptons} is 
identically equal to zero due to the presence of the Levi-Civita symbol in the vertex of $XWW$ interaction.
The amplitude of
the lepton pair production  in the CS boson decays via diagram \textit{a}  in   Fig.\ref{fig:decay_leptons} is given by
\begin{equation}\label{MfiWW}
    M^{WW}_{fi}=\frac{g^2}{2}\bar{\ell}(p') \hat P_R I^{\bar\mu}_W \hat P_L \ell(-p) \epsilon^{\lambda_X}_{\bar{\mu}},
\end{equation}
where $g$ is $SU(2)_L$ coupling, $P_{L(R)}=(1-(+)\gamma^5)/2$,
\begin{equation}\label{Iee}
   I^{\bar\mu}_W=\int\frac{d^4k}{(2\pi)^4}\gamma^{\mu}G_\nu(-p-k)D^{W}_{\mu{\bar{\rho}}}(k+q)\left[c_{\omega}^{\ast}(k+q)_{\bar{\lambda}}+c_{\omega}k_{\bar{\lambda}}\right]D^W_{\bar{\nu}\nu}(k)\gamma^{\nu}\epsilon^{{\bar{\mu}}{\bar{\nu}}{\bar{\lambda}}{\bar{\rho}}}
\end{equation}
and
\begin{equation}\label{DG}
G_f(p)=\frac{m_f+\not p}{m_f^2-p^2},\quad D_{\mu\nu}^V(k)=\frac{g_{\mu\nu}-\frac{k_\mu k_\nu}{M_V^2}}{M_V^2-k^2}
\end{equation}
are propagators for vector field $V$ in unitary gauge and for fermion $f$.

After computation using  the technique of $\alpha$ (Schwinger) representation, see e.g. \cite{Bogolyubov:1983gp}, one can get similarly to \cite{Borysenkova:2021ydf}:
\begin{multline}\label{unitee}
    I^{\bar\mu}_W=
   \hat\Lambda_0^{\nu\ell W} \left\{ \vphantom{\frac12} 
   \gamma_{\bar\rho} (\not{\! \mathcal{P}}-\not{\!p})\gamma_{\bar\nu}\left\{\Theta_W^1(q-2\mathcal{P})_{\bar\lambda}-i\Theta^2_W q_{\bar\lambda} \right\}+\right. \\ +
 \left.  \frac{q_{\bar\lambda}}{M_W^2}
      \mathcal{P}_{\bar\nu}\gamma_{\bar\rho}\left[c_w \left\{2(p\mathcal{P})+\not{\! q\,}\not{\! \mathcal{P}}-\not{\!q\,}\not{\! p}\right\}-2\Theta_W^1  \mathcal{P}^2
      -2i\Theta_W^2  \not{\! p}\not{\! \mathcal{P}}
      \right] \right\} \epsilon^{{\bar{\mu}}{\bar{\nu}}{\bar{\lambda}}{\bar{\rho}}} +  \\
      + \hat\Lambda_1^{\nu\ell W}\left\{ (-i\Theta_W^1)\gamma_{\bar{\rho}}\gamma_{\bar\lambda}\gamma_{\bar{\nu}}+ \frac{q_{\bar\lambda}}{M_W^2} \left[
      i \frac{c_w}2  \gamma_{\bar\rho}{\not{\!q\,}}\gamma_{\bar\nu} -6i\Theta_W^1 \gamma_{\bar\rho}\mathcal{P}_{\bar\nu}+\Theta_W^2 \gamma_{\bar\rho}\not{\!p}\gamma_{\bar\nu}\right] \right\}
      \epsilon^{{\bar{\mu}}{\bar{\nu}}{\bar{\lambda}}{\bar{\rho}}}.
\end{multline}
where
\begin{equation}\label{curlP}
   \mathcal{P} = xp+yq 
\end{equation}
    and $\hat \Lambda_0^{fgV}$, $\hat \Lambda_1^{fgV}$ are integral operators acting on  some function:
\begin{align}
    & \hat \Lambda_0^{fgV}u(x,y) =\frac{i\pi^2}{(2\pi)^4} \int_0^{1}\!\!dx\int_0^{1-x}\!\!\!dy \frac{u(x,y)}{D^{fgV}(x,y)},\label{Lambda0nu} \\
     & \hat \Lambda_1^{fgV} u(x,y) = \frac{-\pi^2}{(2\pi)^4} \int_0^{1}\!\!dx\int_0^{1-x}\!\!\!dy\,  u(x,y)\ln \frac{\Lambda^2 x}{D^{fgV}(x,y)} ,\label{Lambda1nu} \\
     & D^{fgV}(x,y)=x m_f^2-x(1-x)m_g^2 +(1-x)M_V^2- y(1-x-y)M_X^2,\label{Dmnu}
\end{align}
$\Lambda$ is some constant with dimension of mass (it should be put to infinity in the final result, $\Lambda\rightarrow \infty$). So, the divergent part of the loop diagram is hidden in the operator $\hat\Lambda_1^{fgV}$.

\subsection{Production via $XZZ$ interaction}

Lepton pair production via $XZZ$ interaction is described by the diagrams \textit{b} and \textit{f} in  Fig.\ref{fig:decay_leptons}. 
The amplitude of
the lepton pair production  via diagram \textit{f}  in   Fig.\ref{fig:decay_leptons} is 
identically equal to zero due to the presence of the Levi-Civita symbol in the vertex of $XZZ$ interaction.
The amplitude of
the lepton pair production  via diagram \textit{b}  in   Fig.\ref{fig:decay_leptons} is given by
\begin{equation}\label{MfiZZdec}
    M^{ZZ}_{fi}=\frac{g^2}{4\cos^2\theta_W} \,\bar{\ell}(p')\,\, \hat{\overline{P}}_Z I_Z^{\bar\mu} \hat P_Z\, \ell(-p) \epsilon^{\lambda_X}_{\bar{\mu}},
\end{equation}
where 
\begin{equation}\label{PSdec}
    \hat P_Z=t_3^\ell(1-\gamma^5)-2q_\ell \sin^2\theta_W,\qquad  \hat{\overline{P}}_Z =t_3^\ell(1+\gamma^5)-2q_\ell \sin^2\theta_W
\end{equation}
and
\begin{equation}\label{IZdec}
   I_Z^{\bar\mu}=\int\frac{d^4k}{(2\pi)^4}\gamma^{\mu}G_\ell(-p-k)D^Z_{\mu{\bar{\rho}}}(k+q)c_Z (2k+q)_{\bar{\lambda}} D^Z_{\bar{\nu}\nu}(k)\gamma^{\nu}\epsilon^{{\bar{\mu}}{\bar{\nu}}{\bar{\lambda}}{\bar{\rho}}},
\end{equation}
where $G_f(p)$ and $D_{\mu\nu}^V$ were defined in \eqref{DG}, $\theta_W$ is Weinberg angle, $q_f$ is electric charge of lepton $\ell$ in the unites of proton charge and $t_3^{\ell}$   is the third component of the weak isospin ($+1/2$ for neutrinos and $-1/2$ for electrically charged leptons). 

One can get
\begin{multline}\label{unitZZdec}
    I^{\bar\mu}_Z=m_\ell c_Z\hat\Lambda_0^{\ell \ell Z}\left\{\gamma_{\bar\rho} \gamma_{\bar\nu}(q\,-2\mathcal{P})_{\bar\lambda}+\frac{q\,_{\bar\lambda}}{M_Z^2}\mathcal{P}_{\bar\nu}\not{\! q\,} \gamma_{\bar\rho}\right\}\epsilon^{{\bar{\mu}}{\bar{\nu}}{\bar{\lambda}}{\bar{\rho}}} +\\
  + c_Z\hat\Lambda_0^{\ell \ell Z} \left\{ \vphantom{\frac12} 
   \gamma_{\bar\rho} (\not{\! \mathcal{P}}-\not{\!p})\gamma_{\bar\nu}(q-2\mathcal{P})_{\bar\lambda}  +  \frac{q_{\bar\lambda}}{M_Z^2}
      \mathcal{P}_{\bar\nu}\gamma_{\bar\rho} \left(-2  \mathcal{P}^2+2(p\mathcal{P})+\not{\! q\,}\not{\! \mathcal{P}}-\not{\!q\,}\not{\! p}\right)
       \right\} \epsilon^{{\bar{\mu}}{\bar{\nu}}{\bar{\lambda}}{\bar{\rho}}} +  \\
      + i c_Z \hat\Lambda_1^{\ell \ell Z} \left\{ -\gamma_{\bar{\rho}}\gamma_{\bar\lambda}\gamma_{\bar{\nu}}+ \frac{q_{\bar\lambda}}{M_Z^2} \left[
       \frac{ \gamma_{\bar\rho}{\not{\!q}}\gamma_{\bar\nu} }2   -6 \gamma_{\bar\rho}\mathcal{P}_{\bar\nu}\right] \right\}
      \epsilon^{{\bar{\mu}}{\bar{\nu}}{\bar{\lambda}}{\bar{\rho}}},
\end{multline}
where $\hat\Lambda_0^{\ell \ell Z}$, $\hat\Lambda_1^{\ell \ell Z}$ are defined in \eqref{Lambda0nu} and \eqref{Lambda1nu}.

\subsection{Production via $XZA$ interaction}

The amplitude of the lepton pair production via $XZA$ interaction is described by the diagrams \textit{c,d} in   Fig.\ref{fig:decay_leptons}
\begin{equation}\label{MfiZA}
    M_{fi}^{AZ}=\frac{g q_f e}{2\cos\theta_W} \,\bar{\ell}(p') I_{ZA}^{\bar\mu}  \ell(-p) \epsilon^{\lambda_X}_{\bar{\mu}},
\end{equation}
where
\begin{multline}\label{IZA_dec}
   I_{ZA}^{\bar\mu}=c_\gamma \int\frac{d^4k}{(2\pi)^4}\left[ \hat{\overline{P}}_Z \gamma^{\mu}G_\ell(-p-k)D^Z_{\mu{\bar{\rho}}}(k+q) k_{\bar\lambda}
                    D^\gamma_{\bar{\nu}\nu}(k)\gamma^{\nu} +\right.\\
 +\left.  \gamma^{\mu}G_\ell(-p-k)D^\gamma_{\mu{\bar{\rho}}}(k+q) (k+ q)_{\bar\lambda}
                    D^Z_{\bar{\nu}\nu}(k)\gamma^{\nu}\hat{P}_Z\right]\epsilon^{{\bar{\mu}}{\bar{\nu}}{\bar{\lambda}}{\bar{\rho}}}=c_\gamma \hat{\overline{P}}_Z I_{ZA1}^{\bar\mu}+c_{\gamma} I_{ZA2}^{\bar\mu}\hat P_Z,
\end{multline}
$\hat P_Z$, $\hat{\overline{P}}_Z$ are defined by \eqref{PSdec},  $G_f(p)$ and $D_{\mu\nu}^V$ were defined in \eqref{DG}.

One can get
\begin{multline}\label{I_ZA1_dec}
    I_{ZA1}^{\bar\mu}
            = 
            - m_f\left[ \hat\Lambda_0^{1,\ell Z\gamma}\left\{\gamma_{\bar\rho}+\frac{q_{\bar\rho}}{M_Z^2} \,(\not{\!\mathcal{P}} -\not{\!q\,})\right\}\gamma_{\bar\nu}\mathcal{P}_{\bar\lambda}+\hat\Lambda_1^{1,\ell Z\gamma}\frac{i}{2}\frac{q_{\bar\rho}}{M_Z^2}\,\gamma_{\bar\lambda}\gamma_{\bar\nu} \right]\epsilon^{{\bar{\mu}}{\bar{\nu}}{\bar{\lambda}}{\bar{\rho}}}+\\
            +
            \hat\Lambda_0^{1,\ell Z\gamma}\left\{\gamma_{\bar\rho}(\not{\!p}-\not{\mathcal{\!P}})\gamma_{\bar\nu}+\frac{q_{\bar\rho}}{M_Z^2}
            \left[ (\not{\mathcal{\!P}}\not{\!p}-\mathcal{P}^2)\gamma_{\bar\nu}
            -\not{\!q}\,(\not{\!p}-\not{\mathcal{\!P}})\gamma_{\bar\nu}\right] \right\} \mathcal{P}_{\bar\lambda}\epsilon^{{\bar{\mu}}{\bar{\nu}}{\bar{\lambda}}{\bar{\rho}}}+\\
            +
           \hat \Lambda_1^{1,\ell Z\gamma}\frac{i}{2}\left\{ -\gamma_{\bar\rho}\gamma_{\bar\lambda}+\frac{q_{\bar\rho}}{M_Z^2}
            \left( \gamma_{\bar\lambda}\not{\!p}+\not{\!q}\, \gamma_{\bar\lambda}-6\mathcal{P}_{\bar\lambda}\right)\right\}\gamma_{\bar\nu}\epsilon^{{\bar{\mu}}{\bar{\nu}}{\bar{\lambda}}{\bar{\rho}}}
\end{multline}
and
\begin{multline}\label{I_ZA2_dec}
    I_{ZA2}^{\bar\mu}
    =  - m_f \left[ \hat\Lambda_0^{2,\ell Z\gamma}\gamma_{\bar\rho} \left\{\gamma_{\bar\nu}(\mathcal{P}_{\bar\lambda}-q_{\bar\lambda})+\frac{q_{\bar\lambda}}{M_Z^2} \not{\!\mathcal{P}}\mathcal{P}_{\bar\nu}\right\} + \hat\Lambda_1^{2,\ell Z\gamma}   \frac{i}{2} \gamma_{\bar\rho}\frac{q_{\bar\lambda}}{M_Z^2}\gamma_{\bar\nu}
     \right]\epsilon^{{\bar{\mu}}{\bar{\nu}}{\bar{\lambda}}{\bar{\rho}}}+\\
     +
     \hat\Lambda_0^{2,\ell Z\gamma}\gamma_{\bar\rho} \left\{(\not{\!p}-\not{\!\mathcal{P}} ) (\mathcal{P}_{\bar\lambda}-q_{\bar\lambda} )\gamma_{\bar\nu}+
     \frac{q_{\bar\lambda}}{M_Z^2} (\not{\!p}\not{\!\mathcal{P}} -\mathcal{P}^2)\mathcal{P}_{\bar\nu} \right\}\epsilon^{{\bar{\mu}}{\bar{\nu}}{\bar{\lambda}}{\bar{\rho}}}+\\
     +
     \hat\Lambda_1^{2,\ell Z\gamma}\frac{i}{2}\gamma_{\bar\rho}\left\{ -\gamma_{\bar\lambda}\gamma_{\bar\nu}+\frac{q_{\bar\lambda}}{M_Z^2}(\not{\!p} \gamma_{\bar\nu}-6 \mathcal{P}_{\bar\nu}) \right\}\epsilon^{{\bar{\mu}}{\bar{\nu}}{\bar{\lambda}}{\bar{\rho}}},
\end{multline}
where $\hat \Lambda_0^{i, \ell Z\gamma}$, $\hat \Lambda_1^{i, \ell Z\gamma}$ ($i=1,2$) are integral operators acting on a some function
\begin{align}
& \hat \Lambda_0^{i, \ell Z\gamma} u(x,y)=   i\frac{\pi^2}{(2\pi)^4}\int_0^{1}dx\int_0^{1-x}\!\!\!\!dy\frac{u(x,y)}{D^{i,\ell Z\gamma}(x,y)},\label{A0_ZA_dec}\\
& \hat \Lambda_1^{i,\ell Z\gamma} u(x,y)= -\frac{\pi^2}{(2\pi)^4} \int_0^{1}dx\int_0^{1-x}dy\, u(x,y)\ln\frac{\Lambda^2 x}{D^{i, \ell Z\gamma}(x,y)}, \label{A1_ZA_dec}
\end{align}
and
\begin{align}
& D^{1,\ell Z\gamma}(x,y)=x^2 m_\ell^2+y M_Z^2-y(1-x-y) M_X^2, \label{D1gamma_dec}\\
& D^{2,\ell Z\gamma}(x,y)=x^2 m_\ell^2+(1-x-y)M_Z^2-y(1-x-y) M_X^2. \label{D2gamma_dec}
\end{align}

    \section{Combining divergent parts of  diagrams in the\\ direct approach}

Let us look only at divergent parts of the diagrams \eqref{MfiWW}, \eqref{MfiZZdec}, \eqref{MfiZA} to find the possible conditions for cancellation of divergences. It is the parts containing corresponding operators $\hat \Lambda_1^j$, see \eqref{Lambda1nu}, \eqref{A1_ZA_dec}.  We can  see, that  different $\hat \Lambda_1^j$ operators contain equal divergent parts
\begin{multline}
   \hat \Lambda_1^{j} =  \frac{-\pi^2}{(2\pi)^4} \int_0^{1}\!\!dx\int_0^{1-x}\!\!\!dy\,\left[ \ln\frac{\Lambda^2 }{M_W^2}+  \ln\frac{M_W^2 x}{D^{j}(x,y)}\right]=
   \frac{-\pi^2}{2(2\pi)^4}  \ln\frac{\Lambda^2}{M_W^2}+\hat \Lambda_1^{j,finite}\\=L+\hat \Lambda_1^{j,finite}.
\end{multline}
We can write the following useful relation for the operator
\begin{equation}
    \hat\Lambda_1^{j} x =L/3+\hat \Lambda_1^{j,finite} x,
\end{equation}
where 
\begin{equation}
   L= \frac{-\pi^2}{2(2\pi)^4} \ln\frac{\Lambda^2}{M_W^2}\rightarrow \infty.
\end{equation}

The part of 
$ M_{fi} =  M_{fi}^{WW}+ M_{fi}^{ZZ}+ M_{fi}^{AZ}$ proportional to divergent quantity  $L$ is given by
\begin{equation}
    M_{fi}^{div} = L\cdot \bar{\ell}(p') I^{{\mu}} \ell(-p) \epsilon^{\lambda_X}_{\mu},
\end{equation}
where
\begin{equation}\label{infsum}
    I^{{\mu}} =  A\, \gamma_{\bar\rho}\gamma_{\bar\lambda}\gamma_{\bar\nu} \epsilon^{{\bar{\mu}}{\bar{\nu}}{\bar{\lambda}}{\bar{\rho}}}  + B\, q_{\bar\lambda}\gamma_{\bar\rho}\not{q}\gamma_{\bar\nu} \epsilon^{{\bar{\mu}}{\bar{\nu}}{\bar{\lambda}}{\bar{\rho}}}  +  C\, q_{\bar\lambda}\gamma_{\bar\rho}\not{p}\gamma_{\bar\nu} \epsilon^{{\bar{\mu}}{\bar{\nu}}{\bar{\lambda}}{\bar{\rho}}} +D\, q_{\bar\rho}\gamma_{\bar\lambda}\gamma_{\bar\nu} \epsilon^{{\bar{\mu}}{\bar{\nu}}{\bar{\lambda}}{\bar{\rho}}} + E\,  q_{\bar\lambda}\gamma_{\bar\rho}p_{\bar\nu} \epsilon^{{\bar{\mu}}{\bar{\nu}}{\bar{\lambda}}{\bar{\rho}}},
\end{equation}
\begin{multline}\label{coefA}
    A =  \gamma^5 \frac{g}{2} \left[ \frac{\Theta_W^1 g}{2} + \frac{c_z  g}{\cos^2\theta_W}  t_3^f \left[ t_3^f - 2 q_f \sin^2\theta_W \right] + \frac{c_{\gamma} q_f e}{\cos\theta_W}  t_3^f \right] -  \\ - \frac{g}{2} \left[ \frac{\Theta_W^1 g}{2} + \frac{c_z g}{\cos^2\theta_W}   \left(  t_3^f \left[ t_3^f - 2 q_f \sin^2\theta_W \right] + 2 q_f^2 \sin^4\theta_W \right) + \frac{c_{\gamma} q_f e}{\cos\theta_W}   \left( t_3^f-2q_f \sin^2\theta_W \right) \right],
\end{multline}
\begin{multline}\label{coefB}
    B = -  \gamma^5 \frac{g}{4 M_W^2} \left[ \frac{c_w g}{2}  + c_z g  t_3^f \left[ t_3^f - 2 q_f \sin^2\theta_W \right] + c_{\gamma} q_f e  \cos\theta_W  t_3^f \right] + \\
    + \frac{g}{4 M_W^2} \left[ \frac{c_w g}{2}  + c_z g   \left(  t_3^f \left[ t_3^f - 2 q_f \sin^2\theta_W \right] + 2 q_f^2 \sin^4\theta_W \right) +c_{\gamma} q_f e  \cos\theta_W  \left( t_3^f-2q_f \sin^2\theta_W \right) \right],
\end{multline}\vspace{-1em}
\begin{equation}\label{coefC}
    C  =  -i\Theta_W^2 \frac{g^2}{4 M_W^2}(1-\gamma^5),\qquad D =  -\frac{g q_f t_3^f e}{2 \cos\theta_W} \frac{m_f}{M_Z^2} c_\gamma \,  \gamma^5,
\end{equation}\vspace{-1em}
\begin{multline}\label{coefD}
    E =  \gamma^5 \frac{g}{M_W^2} \left[ \frac{\Theta_W^1 g}{2}  + c_z g t_3^f \left[ t_3^f - 2 q_f \sin^2\theta_W \right] + c_{\gamma}  q_f e  \cos\theta_W t_3^f \right] - \\ -  \frac{g}{M_W^2} \left[ \frac{\Theta_W^1 g}{2}  + c_z g   \left(  t_3^f \left[ t_3^f - 2 q_f \sin^2\theta_W \right] + 2 q_f^2 \sin^4\theta_W \right) + c_{\gamma} q_f e  \cos\theta_W  \left( t_3^f-2q_f \sin^2\theta_W \right) \right].
\end{multline}
It would be good to have some relationships between coefficients $\Theta_W^1$, $\Theta_W^2$, $c_\gamma$, $c_z$ of the interaction Lagrangian \eqref{Lcs} like \eqref{L7} -- \eqref{L9}, under which the divergent terms of the loop diagrams are eliminated, i.e. the condition $A=B=C=D=E=0$ would be satisfied. But that's not true.
Putting $C=D=0$, one gets $\Theta_W^2 = 0$ and $c_\gamma=0$.
In this case, the system of the other equations $A=B=E=0$ has only a trivial solution $\Theta_W^1=c_z=0$. 

Expressions \eqref{Iee}, \eqref{unitZZdec}, and \eqref{IZA_dec} contain linear divergent integrals. They will
change when the integration variable is shifted by a constant, as when considering the chiral anomaly,  see \cite{Cheng:1984vwu}. 
However, such a change of variables
will cause the value of the integrals to change only by a finite amount.  So, expressions for $A$ -- $E$ will not change.
    
   \section{Conclusions}

 In this paper, we considered the extension of the Standard Model with the Chern-Simons
type interaction with a new massive vector particle -- Chern-Simons (CS) boson. The Lagrangians of such interaction \eqref{L1}, \eqref{L2} contain operators of dimension 6, but after spontaneous symmetry breaking of the Higgs field, these Lagrangians generate operators of dimension 4 among operators of higher dimensions. Limiting ourselves to considering only the three-particle dimension 4 (possibly renormalizable) interaction of the Chern-Simons bosons with vector bosons of the SM \eqref{Lcs}, we considered the
effective loop interaction of a new vector boson with leptons. 

 As was shown in  \cite{Dror:2017ehi,Dror:2017nsg,Borysenkova:2021ydf} the effective loop interaction (with only $W$-bosons in the loop) of the CS bosons with fermions of different flavours (quarks) does not contain divergences, but interactions with fermions of the same flavours suffer from divergences. But the initial interaction Lagrangian \eqref{Lcs} has no direct interaction of the CS boson with fermions, so we cannot use counterterms to eliminate the divergences.
Therefore, the interaction of the CS bosons with vector bosons of the SM \eqref{Lcs} is self-consistent only if divergences in the effective CS boson's interaction with fermions of the same flavours will be eliminated, accounting for all appropriate loop diagrams.

Taking interaction of the CS bosons with vector fields of the SM in the form \eqref{Lcs}, we considered loop diagrams of the CS boson interaction with leptons that include all possible three-particle vertices, see Fig.\ref{fig:decay_leptons}. We concluded that for the computation of loop diagrams in the unitary gauge, we can not eliminate the divergences in the effective interaction of the Chern-Simons bosons with fermions of the same flavours (leptons), except for the trivial case of the absence of interaction of the Chern-Simons boson with SM particles ($c_w=c_\gamma=c_z=0$), see \eqref{Lcs}. For definitive conclusions, the problem requires a more detailed consideration within the framework of non-unitary gauge. 

\section{Acknowledgments}

The work of V.G., I.H. and O.Kh. was supported by the National Research Foundation of Ukraine under project No. 2023.03/0149.
The authors are grateful to  Oleg Ruchayskiy and Eduard Gorbar for fruitful discussions and helpful comments.

\bibliographystyle{JHEP}
\bibliography{Bibl}

\end{document}